\documentclass[aip,amsmath,amssymb,reprint,onecolumn]{revtex4-1}

\usepackage{latexsym}
\usepackage{amssymb, amsfonts, amsmath}
\usepackage{amscd}
\usepackage[all]{xy}
\usepackage{amsthm}
\usepackage{enumerate}   
\usepackage{bm}
\usepackage{empheq}
\usepackage{hyperref}
\usepackage{graphicx}		
\usepackage{cleveref}

\newtheorem{thm}{Theorem} 
\newtheorem{prop}{Proposition}
\newtheorem{lem}{Lemma}

\theoremstyle{definition}

\newtheorem{expl}{Example}

\newtheorem*{slogan*}{SLOGAN}

\newtheorem{rem}{Remark}

\usepackage{colortbl}

\newcommand{\wk}[1]{{\mathrm{w},#1}}
\newcommand{\bmxw}{\bm{x}^{\rm w}}
\newcommand{\bmv}{{\bm v}}
\newcommand{\xw}{x^{\rm w}}
\newcommand{\dxw}{\dot{x}^{\rm w}}
\newcommand{\ddxw}{\ddot{x}^{\rm w}}
\newcommand{\bmyw}{\bm{y}^{\rm w}}
\newcommand{\yw}{y^{\rm w}}
\newcommand{\dyw}{\dot{y}^{\rm w}}

\newcommand{\zw}{z^{\rm w}}

\newcommand{\R}{\mathbb{R}}

\DeclareMathOperator{\ci}{\mathrm{i}}

\begin{document}

\title{Error analysis of the complex-valued Kuramoto model}

\author{Robin Delabays}
\email{robin.delabays@hevs.ch}
\affiliation{University of Applied Sciences of Western Switzerland (HES-SO), CH-1950 Sion, Switzerland}

\author{Longfei Gao}
\email{longfei.gao@anl.gov}
\affiliation{Argonne National Laboratory}

\author{Tung T. Nguyen}
\email{tung.nguyen@elmhurst.edu}
\affiliation{Department of Mathematics, Elmhurst University}


\begin{abstract}
The complex-valued Kuramoto model was recently introduced as an approximation of the networked Kuramoto model. 
Combined with a windowed approach, the complex-valued Kuramoto model has empirically been shown to be a very good approximation of the original Kuramoto model, with the strength of possessing a closed-form solution.  We propose a rigorous estimate of the error made by approximating the Kuramoto model with its complex-valued counterpart.  In particular, we show that the error grows at most linearly with the window size of the windowed approach for fixed simulation duration. As a consequence, we show that the global error of the Kuramoto order parameters of these two models is also at most linear with respect to the window size. 
\end{abstract}

\maketitle

\section{Introduction}
Since its introduction in 1975,~\cite{kuramoto1975self} the Kuramoto model has established itself as the canonical system of ordinary differential equations (ODEs) to describe the synchronization of coupled oscillators.~\cite{strogatz2000kuramoto} 
Originally motivated as a first-order approximation for a group of limit cycle oscillators, the Kuramoto model soon became a research topic in its own right.  Indeed, the richness of the model, even in its simplest variants, justifies thorough investigations.
The Kuramoto model has been used to describe synchronization phenomena in various areas, including biological systems, power networks, chemical oscillators, and neural networks (see Refs.~\onlinecite{strogatz2000kuramoto, acebron2005kuramoto, arenas2008synchronization, dorfler2013synchronization}). 

Despite its relatively simple form, the model is able to capture important collective behaviors such as phase synchronization, clustering, and the transition from incoherence to synchronization.
In its original version,~\cite{kuramoto1975self} the Kuramoto model considers a population of phase oscillators coupled through a mean-field interaction. 
The main tuning variable in this model is the distribution of the oscillators' natural frequencies. 
Nowadays, variations of the Kuramoto model include higher-order couplings,~\cite{leon2019phase,leon2025theory,muolo2026when} phase frustrations,~\cite{sakaguchi1987local} and adaptive couplings,~\cite{gross2007adaptive} to name but a few examples. 

A major ingredient of the model's richness is undoubtedly the strong nonlinearity introduced by the sinusoidal coupling between phases. 
The flip side of this richness is that, in general, closed-form solutions do not exist for the Kuramoto model. 
Therefore, one has to rely on numerical approximations in order to get a glimpse at the system's behavior. 

Recently, the complex-valued Kuramoto model~\cite{muller2021algebraic} has been introduced in an attempt to circumvent this difficulty.
The strength of the complex-valued Kuramoto model is that it admits a closed-form solution.
Although the original and the complex-valued Kuramoto models are not mathematically equivalent (see \Cref{sec:kuramoto}), the latter has been shown to be a very good approximation to the former when combined with a windowed approach.~\cite{muller2021algebraic, budzinski2021simple, nguyen2023equilibria, budzinski2023analytical} 
However, to this day, no systematic analysis of the error produced by the windowed complex-valued Kuramoto model has been conducted. 

In this manuscript, we provide a rigorous quantification of the discrepancy between the original and the complex-valued models, with respect to both the window size $\tau$ and the simulation duration $T$. 
In particular, we show that for a fixed duration $T$, the discrepancy between the two models is at most linear in the window size $\tau$ (see \Cref{thm:main}). 
Furthermore, using this result, we show that the difference between the Kuramoto order parameters of these two models is also at most linear with respect to $\tau$ (see \Cref{thm:main-2}).

\section{The Kuramoto models}\label{sec:kuramoto}
We first recall the original Kuramoto model and then, building on it, introduce the complex-valued Kuramoto model and the windowed approach, which is the focus of this work.

\subsection{The original Kuramoto model}\label{sec:kuramoto_original}
The Kuramoto model describes the time evolution of a set of $N$ phase oscillators, each of which moves on the unit circle and is parametrized by its phase $\theta\in\mathbb{R}$. 
It is typically formulated as the following system of ODEs 
\begin{align} \label{eq:original_KM}
    \dot{\theta}_i &= \sum_{j=1}^N a_{ij} \sin(\theta_j - \theta_i)\, ,
\end{align}
where $i\in\{1,\ldots,N\}$ and $A=(a_{ij})_{1 \leq i, j \leq N}$ is the (weighted) adjacency matrix with $a_{ij}\in\mathbb{R}$.
Below, we use ${\bm \theta} = (\theta_1, \theta_2, \ldots, \theta_N)^\top \in \mathbb{R}^N$ to denote the vector of all phases in the network. 
Since, in practice, we usually focus on undirected graphs, $A$ is assumed to be a non-zero symmetric matrix with zero diagonal entries. 
Consequently, $A$ is diagonalizable with all its eigenvalues real.
We will see later that the spectrum of $A$ plays an important role in the error analysis of the complex-valued Kuramoto model.

\begin{rem}
We note that the early occurrences of the Kuramoto model actually considered mean-field systems. 
Such a system would correspond to having an all-to-all interaction graph with all weights $a_{ij}$, where $i \neq j$, set to $1/N$.

Moreover, in general, each oscillator has its own natural frequency $\omega_i$ driving its dynamics in addition to the sinusoidal couplings. 
Here, we consider a set of oscillators sharing a common natural frequency $\omega$, which, in this case, can be eliminated from the system by a simple change of variable.

\end{rem}

In the following, we denote the solution of \Cref{eq:original_KM} with initial condition $\bm{\theta}_0\in\mathbb{R}^N$ as ${\bm \theta}(t;{\bm \theta}_0)$, or simply ${\bm \theta}(t)$ when the initial condition is clear from the context.
Furthermore, the time dependence may also be omitted when no confusion arises.

\subsection{The complex-valued model}
In Ref.~\onlinecite{muller2021algebraic}, the following complex-valued ODE system was introduced 
\begin{align} \label{eq:complex_KM}
    \dot{\zeta}_{i} &= \sum_{j=1}^N a_{ij} [\sin(\zeta_j-\zeta_i) - \mathrm{i} \cos(\zeta_j-\zeta_i) ]\, ,
\end{align}
where ${\rm i}$ is the imaginary unit and $\zeta_i\in\mathbb{C}$ is the $i$th complex oscillator.
Subsequent works in Refs.~\onlinecite{budzinski2021simple, nguyen2023equilibria, budzinski2023analytical} utilized this complex-valued Kuramoto model to provide some geometric explanations for several phenomena in the original Kuramoto model including equilibria, transient dynamics, and synchronization. 

The key observation in Ref.~\onlinecite{muller2021algebraic} is that the complex-valued Kuramoto model described by \Cref{eq:complex_KM} admits a closed-form solution. 
More specifically, if we multiply both sides of \Cref{eq:complex_KM} by $\mathrm{i}$ and apply Euler's formula, we have 
\begin{align} \label{eq:complex_original} 
    \ci \dot{\zeta}_i &= \sum_{j=1}^N a_{ij} [\cos(\zeta_j-\zeta_i) + \ci \sin(\zeta_j - \zeta_i)] = \sum_{j=1}^N a_{ij} e^{\ci (\zeta_j - \zeta_i)}\, . 
\end{align}
Multiplying both sides of \Cref{eq:complex_original} by $e^{\ci \zeta_i}$ leads to
\begin{align}  \label{eq:exact_form}
    \dfrac{d}{dt} (e^{\ci \zeta_i}) &= \sum_{j=1}^N a_{ij} e^{\ci \zeta_j}\, ,
\end{align}
which can be rewritten as
\begin{align}\label{eq:linear_form}
    \dot{\bm z} &= A\bm{z}\, ,
\end{align}
by introducing the variables $z_i = e^{\ci \zeta_i}$ and $\bm{z} = (z_1,...,z_N)^\top$. 
\Cref{eq:linear_form} has the closed-form solution 
\begin{align}
    {\bm z}(t) &= e^{At} {\bm z}(0)\, ,
\end{align}
where ${\bm z}(0)$ is the initial condition to be specified later. 

\begin{rem}
    In this work, we focus on the classical Kuramoto model and its complex-valued analogue for clarity of exposition. However, our analysis extends naturally to the Kuramoto model with phase lags discussed in Ref.~\onlinecite{budzinski2021simple}. 
    In particular, Theorems \ref{thm:main} and \ref{thm:main-2} can still be established in this setting, after minor modifications to the proof of \Cref{lem:bound_y}. 
\end{rem}

\subsection{Relation between the original and complex-valued Kuramoto models}
To better appreciate the connection between the original and the complex-valued Kuramoto models, we express each complex phase $\zeta_i$ in terms of its real and imaginary parts $\zeta_i =: x_i+\ci y_i$. 
Using $x_i$ and $y_i$, we can rewrite \Cref{eq:complex_original} as the following system of ODEs
\begin{subequations} \label{eq:complex_Cartesian}
\begin{empheq}[left=\empheqlbrace]{alignat=2}
    \dot{x}_i &=   && \sum_{j=1}^N a_{ij} e^{-(y_j-y_i)}
        \sin(x_j - x_i)\, ; \label{eq:dotx}\\
    \dot{y}_i &= - &&\sum_{j=1}^N a_{ij} e^{-(y_j-y_i)}
        \cos(x_j-x_i)\, . \label{eq:doty}
\end{empheq}
\end{subequations}

Notice that if $y_i=0$ $\forall i\in\{1,\ldots,N\}$, \Cref{eq:dotx} reduces to the original Kuramoto model. 
However, since $y_i$ evolves in time according to \Cref{eq:doty}, it generally drifts away from zero even if $y_i=0$ $\forall i\in\{1,\ldots,N\}$ initially, leading to a discrepancy between the two models. 

In the upcoming discussion, we use $\big(\bm{x}(t;\bm{x}_0,\bm{y}_0),\bm{y}(t;\bm{x}_0,\bm{y}_0)\big)$ to denote the solution of \Cref{eq:complex_Cartesian} with initial condition $(\bm{x}_0,\bm{y}_0)$, or simply $\big(\bm{x}(t),\bm{y}(t)\big)$ when no confusion arises.

\subsection{The windowed approach}\label{sec:windowed_approach}
Let the total observation time $T$ be divided into $n$ windows of equal size $\tau$ (i.e., $T=n\cdot\tau$). 
To mitigate the discrepancy between the two Kuramoto models, we rely on the \emph{windowed approach}, where the original Kuramoto model, i.e., the initial value problem described by \Cref{eq:original_KM}, is split into $n$ small initial value problems, each restricted to one of these windows and approximated by the complex-valued Kuramoto model.
The windowed approach proceeds as follows:
\begin{enumerate}
    \item 
    In the first window, i.e., the time interval $[0,\tau]$, the original Kuramoto model is approximated by the complex-valued Kuramoto model on $[0,\tau]$ with initial conditions $\bm{x}^\wk1_0 = \bm{\theta}_0$ and $\bm{y}^\wk1_0 = \bm{0}$.
    The closed-form solution of this approximate model serves as our approximate solution of the original model in the first window. 
    We denote the approximate solution thus obtained as $\big(\bm{x}^\wk1(t), \bm{y}^\wk1(t)\big) \coloneqq
    \big(\bm{x}(t;{\bm x}^\wk1_0, \bm{0}), \bm{y}(t;{\bm x}^\wk1_0,\bm{0})\big)$. 
    Let $\bm{z}^\wk1(t)$ be the complex-valued vector whose $i$th component $z^\wk1_i$ is defined as $z^\wk1_i = e^{\ci (x^\wk1_i + \ci y^\wk1_i)}$, where $x^\wk1_i$ and $y^\wk1_i$ are the $i$th components of $\bm{x}^\wk1(t)$ and $\bm{y}^\wk1(t)$, respectively.
    From \Cref{eq:linear_form}, we have
    \begin{align}
        \bm{z}^\wk1(t) = e^{At} e^{ \ci \bm{x}^\wk1_0},
        \enskip \forall t\in[0,\tau],
    \end{align}
    where the exponential in $e^{ \ci \bm{x}^\wk1_0}$ is taken componentwise.
    We note here that the relations
    $x^\wk1_i \equiv \arg(z^\wk1_i) \pmod{2\pi}$
    and 
    $e^{-y^\wk1_i} = \big|z^\wk1_i\big|$ hold by the properties of complex numbers.
    
    \item In the second window, i.e., the time interval $[\tau,2\tau]$, the original Kuramoto model is approximated by the complex-valued Kuramoto model on $[0,\tau]$ with initial conditions ${\bm x}^\wk2_0 = \bm{x}^\wk1(\tau)$ and $\bm{y}^\wk2_0 = \bm{0}$. 
    In other words, the approximation ${\bm x}^\wk1(\tau)$ at the end of the first window is retained as the initial condition in the second window, whereas the approximation ${\bm y}^\wk1(\tau)$ is replaced by $\bm 0$.
    As in the first window, we have from \Cref{eq:linear_form}
    \begin{align}\label{eq:solution_2nd_window}
        \bm{z}^\wk2(t) = e^{At} e^{\ci \bm{x}^\wk2_0}
        ,\enskip \forall t\in[0,\tau]. 
    \end{align}

    \item The same procedure is repeated for the subsequent windows, leading to 
    \begin{align}\label{eq:solution_all_windows}
        \bm{z}^\wk{k+1}(t) = e^{At} e^{\ci \bm{x}^\wk{k+1}_0}
        ,\enskip \forall t\in[0,\tau],
    \end{align}
    for $k=1,\ldots,n-1$, where $\bm{x}^\wk{k+1}_0 = \bm{x}^\wk{k}(\tau)$.
\end{enumerate}

We note that, in the above procedure, the windowed
solution $\bm{x}^\wk{k+1}(t)$, $t\in[0,\tau]$,
approximates $\bm{\theta}(t+k\tau)$, i.e., the solution
of the original Kuramoto model on the $(k+1)$th window,
for $k=0,\ldots,n-1$.

\subsection{Numerical example}
We illustrate our results on a toy example of networked Kuramoto oscillators. 
We consider a system composed of $N=20$ oscillators, coupled according to a random Erd\H{o}s-R\'enyi graph with coupling strength $1/N$. 
In \Cref{fig:kuramoto-comparison} we approximate the true Kuramoto dynamics with the windowed complex Kuramoto model, using three different window sizes. 
In \Cref{fig:one_window} and \Cref{fig:global_error} the local and global error of the approximation as functions of the window size.

All simulations were carried out using the code available at Ref.~\onlinecite{delabays_gao_nguyen_kuramoto_code_2026}.

\section{Error estimates}

\subsection{Quantities of interest and main results}
Consider the original and the complex-valued Kuramoto models, given by \Cref{eq:original_KM,eq:complex_KM}, respectively, with the same interaction graph $A$. 
The discrepancy between the solution of the original model and its approximation via the windowed approach at the end of the $k$th window can be quantified by
\begin{align}\label{eq:error}
    E_{k} := \|{\bm \theta}(k \tau) - \bm{x}^\wk{k}(\tau)\|_{\infty} = \max_{1 \leq i \leq N} |\theta_i(k \tau) - x^\wk{k}_i(\tau)|\, .
\end{align}
Since the $\infty$-norm is used here, $E_{k}$ is determined by the oscillator with the largest approximation error.

In addition to the individual trajectories, the scalar Kuramoto order parameter defined below is a key indicator of the network's collective behavior, as it captures the degree of synchrony among the oscillators:
\begin{align}\label{eq:order_parameter_R}
    r &= \frac{1}{N} \left|\sum_{i=1}^N e^{ \ci \theta_i} \right|\, .
\end{align}
The dependence of $r$ and $\theta_i$ on time $t$ is
omitted for notational clarity.
It is easy to see that $r$ ranges between $0$ and $1$, with $r=1$ indicating perfect synchrony when all oscillators share the same phase and $r=0$ indicating complete incoherence when the phases of the oscillators cancel each other out.
In the following, we use $R_k$ to denote the value of $r$ at the end of the $k$th window, i.e., $R_k = r(k\tau)$.

In the windowed approach, the order parameter $r$ is approximated by 
\begin{align}\label{eq:order_parameter_R_complex_KM}
    r^\wk{k} &= \frac{1}{N} \left|\sum_{i=1}^N e^{ \ci x^\wk{k}_i} \right| 
\end{align}
in the $k$th window.
Similarly, we use $R^{\rm w}_k$ to denote the value of $r^\wk{k}$ at the end of the $k$th window, i.e.,
$R^{\rm w}_k = r^\wk{k}(\tau)$.
The discrepancy $\left|R_{k}-R^{\rm w}_{k}\right|$ is a simple measure that summarizes the difference in the collective behavior of the two systems at the end of the $k$th window and is therefore a quantity of interest in our subsequent analysis.

\begin{figure*}[!htbp]
    \centering
    \includegraphics[width=\textwidth]{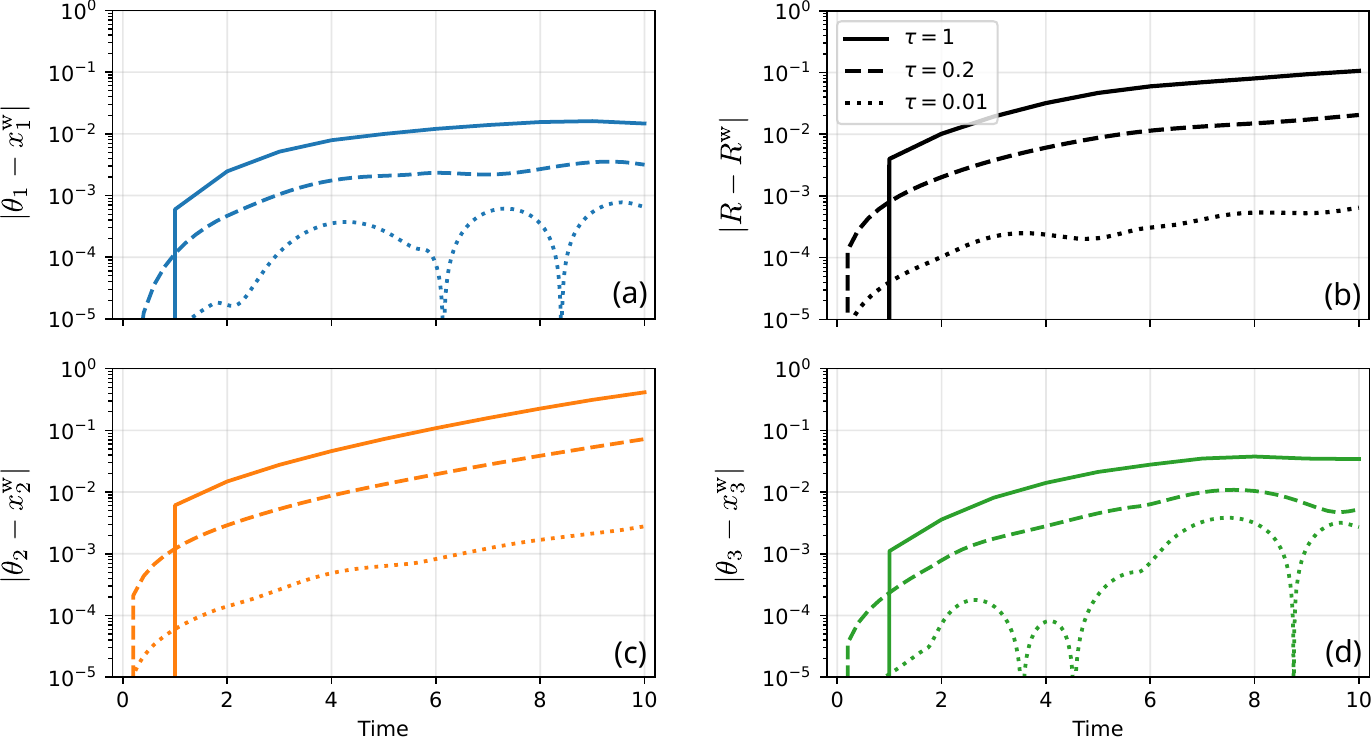}
    \caption{Errors in the approximate solutions from the windowed approach with three different window sizes, $\tau=0.01$ (dotted), $\tau=0.2$ (dashed), and $\tau=1.0$ (plain).
    The interaction graph is a random Erd\H{o}s-R\'enyi graph with $N=20$ vertices and coupling strength $1/N$. 
    Panels (a), (c), and (d) show the errors in the trajectories of three arbitrarily selected nodes. 
    Panel (b) shows the errors in the order parameter.}
    \label{fig:kuramoto-comparison}
\end{figure*}

One can expect that a larger window size would lead to a larger approximation error in the solution from the windowed approach.
This can be observed numerically, as illustrated in \Cref{fig:kuramoto-comparison}, where the discrepancies in the oscillator trajectories and the order parameter are shown in the left and right panels, respectively, for three different window sizes. 

We aim to rigorously quantify these errors in the analysis below. 
Our first result establishes the dependence of $E_{k}$ on $\tau$, which is summarized in \Cref{thm:main} and proved later.

\begin{thm} \label{thm:main}
    The global error $E_{k}$ is at most linear with respect to $\tau$; namely, $E_{k} = O(\tau).$ 
\end{thm}

With \Cref{thm:main}, the following result on $|R_{k}-R^{\rm w}_{k}|$ follows immediately from the relation
$\left|R_{k}-R^{\rm w}_{k}\right| \leq E_{k}$, which is established in Appendix~\ref{app:estimate}.

\begin{thm} \label{thm:main-2}
    The order parameter error $|R_{k}-R^{\rm w}_{k}|$ is also at most linear with respect to $\tau$.
\end{thm}

To facilitate the subsequent analysis, we introduce an auxiliary notation ${\bm \theta}^\wk{k}(t):= \bm{\theta}(t;\bm{x}^\wk{k}_0)$  for $t\in[0,\tau]$, which denotes the solution of the original Kuramoto model restricted to the $k$th window, but started from the initial condition ${\bm x}^\wk{k}_0$ at the beginning of this window instead of ${\bm \theta}((k-1)\tau)$.
We first note that the global error at the end of the $k$th window can be split into two parts as follows:
\begin{align}\label{eq:error_split}
 E_{k} 
 &= \left\|\bm{x}^\wk{k}(\tau) - \bm{\theta}(k\tau)
 \right\|_{\infty} 
 \leq \underbrace{\left\|\bm{x}^\wk{k}(\tau) - \bm{\theta}^\wk{k}(\tau)\right\|_{\infty}}_{\text{local error}}
 + \underbrace{\left\|\bm{\theta}^\wk{k}(\tau) - \bm{\theta}(k\tau) \right\|_{\infty}}_{\text{initial-state error}}\, .
\end{align}
The first term in \Cref{eq:error_split} describes the local error committed in the $k$th window due to the discrepancy between the original and the complex-valued Kuramoto models; the second term accounts for errors committed in the previous windows, which are carried over in the initial condition for the $k$th window.
They are analyzed in \Cref{sec:local_error,sec:initial_state_error}, respectively. 

From the definitions of the dynamics of $\bmxw$ and $\bm{\theta}^{\rm w}$, one verifies that the local error grows at most quadratically in time. 
In \Cref{thm:main-error-1}, we prove that a global quadratic bound exists. 
This global bound is obtained by bounding the second time derivatives of $\bmxw$ and $\bm{\theta}^{\rm w}$ (\Cref{prop:bound_theta} and \Cref{prop:bound_x_t_w}) with values that are only graph-dependent.

The initial-state error is bounded by a direct application of Gr\"onwall's inequality of the difference between the two trajectories (\Cref{thm:stability_error}). 

\subsection{Local error}\label{sec:local_error}
Below, we analyze the local error committed in one window and quantify the difference between $\bmxw(t) := \bm{x}(t;\bm{\xi}, \bm{0})$ and $\bm{\theta}^{\rm w}(t) := \bm{\theta}(t;\bm{\xi})$ for an arbitrary initial condition $\bm{\xi} \in \R^N$.
Here, $\bm{\theta}^{\rm w}$ satisfies \Cref{eq:original_KM} on interval $[0,\tau]$ with initial condition $\bm{\xi}$, whereas $(\bmxw, \bmyw)$ satisfy \Cref{eq:complex_Cartesian} on interval $[0,\tau]$ with initial condition $(\bm{\xi},\bm{0})$.

Before presenting the rigorous proof, let us explain the intuition behind our approach.
The dynamics of $\bm{\theta}^{\rm w}$ and $\bmxw$ start with the same initial condition $\bm{\xi}$.
By comparing \Cref{eq:original_KM,eq:complex_Cartesian}, it is easy to see that at time $t=0$, the first derivatives of 
$\bm{\theta}^{\rm w}$ and $\bmxw$  are the same as well.
Therefore, the Taylor expansions of the components of $\bm{\theta}^{\rm w}$ around $t=0$ and $\bmxw$ have the same zeroth- and first-order terms.
Consequently, for any $\bm{\xi}$, there exists a constant $C_{\xi}$ such that
\begin{align}
    \|\bm{\theta}^{\rm w} (t) - \bmxw (t)\|_\infty \leq C_{\xi} t^2,
    \enskip \forall t \in [0,\tau].
\end{align}
However, this estimate is not sufficient for the global error analysis because the constant $C_{\xi}$ may depend on the initial condition $\bm{\xi}$, which changes from window to window.
The key point of the following theorem is that this constant can be chosen independently of the initial condition $\bm{\xi}$.

\begin{thm} \label{thm:main-error-1}
There exist explicit constants $\tau_0>0$ and $C_A>0$, which depend only on the adjacency matrix $A$, such that for any chosen window size $\tau\leq\tau_0$, the following relation regarding the components of $\bmxw$ and $\bm{\theta}^{\rm w}$ holds for an arbitrary initial condition $\bm{\xi}$:
\begin{align}\label{eq:local_error_bound}
    |\theta^{\rm w}_i (t) - \xw_i (t)|\leq C_At^2,
    \enskip \forall t \in [0,\tau],
    \enskip i = 1,\ldots, N.
\end{align}
In other words, the local error is $O(t^2)$ regardless of the initial condition $\bm{\xi}$.
\end{thm}

As noted above, $\bm{\theta}^{\rm w}$ and $\bmxw$ share the zeroth- and first-order terms in their Taylor expansions.
Therefore, to prove \Cref{thm:main-error-1}, it is sufficient to show that their second derivatives are both uniformly bounded, which will be established in Propositions \ref{prop:bound_theta} and \ref{prop:bound_x_t_w}, respectively.

\begin{prop} \label{prop:bound_theta}
Let $\Delta$ be the largest weighted degree of the graph with adjacency matrix $A=(a_{ij})$; namely 
\begin{equation} \label{eq:Delta_infinity}
\Delta = \max_i \left(\sum_{j=1}^N |a_{ij}| \right)\, .    
\end{equation}
Then, for $\bm{\theta}$ satisfying (\ref{eq:original_KM}) with arbitrary initial condition, we have $|\ddot\theta_i| \leq2 \Delta^2$ for each $1 \leq i \leq N$. 
\end{prop}
\begin{proof}
First, we notice that by  the triangle inequality and the fact that $|\sin(\theta_j-\theta_i)| \leq 1$, we have 
\begin{align*}
    |\dot \theta_i | &= \left|\sum_{j=1}^N a_{ij} \sin(\theta_j-\theta_i) \right| \leq \sum_{j=1}^N |a_{ij}| \leq \Delta\, .
\end{align*}
By taking the derivative of $\dot{\theta}_i$ and using the triangle inequality again, we have 
\begin{align*}
    |\ddot{\theta}_i| = \left| \sum_{j=1}^N a_{ij} \cos(\theta_j-\theta_i) (\dot \theta_j - \dot \theta_i) \right| \leq \sum_{j=1}^N |a_{ij}|(|\dot{\theta}_j|+ |\dot{\theta}_i|) \leq 2 \Delta \sum_{j=1}^N |a_{ij}| \leq 2 \Delta^2\, .
\end{align*}
\end{proof}

Before showing that $\ddot{\bm x}^{\rm w}$ is uniformly bounded, let us recall some properties of the adjacency matrix. 
Recall that we focus on undirected graphs and that the non-zero adjacency matrix $A$ is real and symmetric with zero diagonal entries. 
It admits eigen-decomposition $A = V \Lambda V^\top$, where $\Lambda \in \mathbb R^{N \times N}$ is diagonal with diagonal entries being the eigenvalues $\lambda_1, \ldots, \lambda_N$ and the columns of $V \in \mathbb R^{N \times N}$, denoted by $\bmv_1, \bmv_2, \ldots, \bmv_N$, are the corresponding eigenvectors that form an orthonormal basis. 
Written in matrix notations, we have $V^\top V = VV^\top = I$.
Since $A$ has zero diagonal entries, we have $\operatorname{tr}(A)=\sum_{j=1}^N \lambda_j=0$. 
Ordering the eigenvalues such that $\lambda_1 \geq \lambda_2 \geq \ldots \geq \lambda_N$, we can conclude that $\lambda_1 > 0$ and $\lambda_N < 0$. 

A bound on $|e^{\lambda_j t}-1|$, $j = 1, \ldots, N$, is used in the proof below, which can be obtained by restricting the range of $t$. 
We aim to have $|e^{\lambda_j t}-1| \leq \frac{1}{2\sqrt N}$, the reason for this will become apparent in the Lemma below.
This bound is guaranteed by 
$e^{\lambda_1 t} \leq 1 + \frac{1}{2\sqrt N}$ 
and
$e^{\lambda_N t} \geq 1 - \frac{1}{2\sqrt N}$,
given $\lambda_1 \geq \lambda_j \geq \lambda_N$.
In other words, denoting
\begin{equation}\label{eq:tau_0}
\tau_0 = \min\left\{ 
\frac{\log\left(1+\frac{1}{2\sqrt N}\right)}{\lambda_1}, \frac{\log\left(1-\frac{1}{2\sqrt N}\right)}{\lambda_N} \right\},
\end{equation}
for any window size $\tau \leq \tau_0$, we have $|e^{\lambda_j t}-1| \leq \frac{1}{2\sqrt N}$ for $t \in [0,\tau]$ and $1 \leq j \leq N$.

For the remainder of the local error analysis, we will assume that the window size is chosen as such.
We note here that although this specification of $\tau_0$ is sufficient for the proof given below, it may be overly stringent in practice.

\begin{lem} \label{lem:bound_y}
    There exists a constant $\varepsilon$ independent of $\tau$ and $\bm{\xi}$ such that $|\yw_i (t)| \leq \varepsilon $ for $t \in [0,\tau]$ and $1 \leq i \leq N$.
\end{lem}

\begin{proof}
Recall that 
$z^{\rm w}_i = e^{\ci (x^{\rm w}_i + \ci y^{\rm w}_i)}$ and 
$|z^{\rm w}_i| = e^{-y^{\rm w}_i}$.
Bounding $|z^{\rm w}_i|$ on both sides is therefore sufficient to establish a bound on $|y^{\rm w}_i|$.
From $\boldsymbol{z}^{\rm w}(t) = e^{At}e^{\ci\bm{\xi}}$, the initial state of ${\bm z}^{\rm w}$, i.e., ${\bm z}^{\rm w}(0) = e^{\ci\bm{\xi}}$, can be expressed as 
\begin{equation}\label{eq:transformation_z0}
e^{\ci\bm{\xi}} 
= \sum_{j=1}^N c_j \bmv_j
\end{equation}
using the eigen-basis, which in turn leads to 
\begin{align*}
    \bm{z}^{\rm w}(t) = \sum_{j=1}^N c_j e^{At} \bmv_j
    = \sum_{j=1}^N c_j e^{\lambda_j t} \bmv_j
    = e^{\ci\bm{\xi}} + \sum_{j=1}^N c_j \left(e^{\lambda_j t} - 1\right) \bmv_j.  
\end{align*}

Written componentwise, we have 
\begin{equation}\label{eq:z^w_i_split}
z^{\rm w}_i(t)
= e^{\ci{\xi}_i} + \sum_{j=1}^N c_j \left(e^{\lambda_j t} - 1\right) v_{ij}.
\end{equation}
Recall that $VV^\top =I$, which leads to 
$\left(\sum_{j=1}^N v_{ij}^2\right)^\frac{1}{2} = 1$.
Moreover, \Cref{eq:transformation_z0} can be written in matrix notation as $e^{\ci\bm{\xi}} = V \bm{c}$, which gives 
$\|e^{\ci\bm{\xi}}\|_2 = \|V \bm{c}\|_2 = \|\bm{c}\|_2 
\coloneqq \left(\sum_{j=1}^N |c_{j}|^2\right)^\frac{1}{2}$ following $V^\top V = I$.
We therefore have $\left(\sum_{j=1}^N |c_{j}|^2\right)^\frac{1}{2} = \sqrt N$ since $|e^{\ci{\xi}_i}| = 1$.
Using the bound on $\left| \big(e^{\lambda_j t}-1\big) \right|$ established earlier and the Cauchy--Schwarz inequality, we arrive at 
\begin{align*}
    \Big|\sum_{j=1}^N \big(e^{\lambda_j t}-1\big) c_j v_{ij}\Big| &\leq \frac{1}{2\sqrt N}\, \Big(\sum_{j=1}^N |c_{j}|^2\Big)^\frac{1}{2}\Big(\sum_{j=1}^N v_{ij}^2\Big)^{1/2} = \frac{1}{2} ,
\end{align*}
which leads to $\tfrac12 \leq |\zw_i(t)| \leq \tfrac32$ following \Cref{eq:z^w_i_split}.
This eventually gives $\log(\frac{2}{3}) \leq y^{\rm w}_i (t) \leq \log(2)$ and $\left|y^{\rm w}_i (t) \right| \leq \log(2) =: \varepsilon$, which is the uniform bound we sought after.
\end{proof}

\begin{prop} \label{prop:bound_x_t_w}
Let $\Delta$ be as given in (\ref{eq:Delta_infinity}) and 
$\varepsilon$ be as given in \Cref{lem:bound_y}. 
For $i = 1, \ldots, N$,  we have 
$|\dyw_i| \leq e^{2 \varepsilon} \Delta$, 
$|\dxw_i| \leq e^{2 \varepsilon} \Delta$, and
$|\ddxw_i| \leq 4 e^{4 \varepsilon} \Delta^2$ for $t\in [0,\tau].$
\end{prop}

\begin{proof}
Recall that $(\bmxw, \bmyw)$ satisfy \Cref{eq:complex_Cartesian} on interval $[0,\tau]$ with initial condition $(\bm{\xi},\bm{0})$.
We have 
\begin{align*}
    \dyw_{i} = - \sum_{j=1}^N a_{ij} e^{-(\yw_j-\yw_i)} \cos(\xw_j-\xw_i)\, .
\end{align*}
From \Cref{lem:bound_y}, we have $|\yw_i| \leq \varepsilon$ and therefore $e^{-(\yw_j-\yw_i)} \leq e^{2\varepsilon}$. 
Noticing that $|\cos(\xw_j-\xw_i)| \leq 1$ and $\sum_{j=1}^N |a_{ij}| \leq \Delta$, we have $|\dyw_i| \leq e^{2 \varepsilon} \Delta$ by simply applying the triangle inequality.
Following a similar approach, one can easily arrive at $|\dxw_i| \leq e^{2 \varepsilon} \Delta$ as well, where 
\begin{align*}
    \dxw_i = \sum_{j=1}^N a_{ij} e^{-(\yw_j-\yw_i)} \sin(\xw_j - \xw_i)\, .
\end{align*}
To obtain the bound on $|\ddxw_i|$, we take the derivative on both sides of the above equation, arriving at $\ddxw_i = A_i-B_i, $ where 
\begin{align*}
    A_i = \sum_{j=1}^N a_{ij} e^{-(\yw_j-\yw_i)} \cos(\xw_j - \xw_i) (\dxw_j - \dxw_i)\, ,
\end{align*}
and 
\begin{align*}
    B_i = \sum_{j=1}^N a_{ij} e^{-(\yw_j-\yw_i)} \sin(\xw_j - \xw_i) (\dyw_j - \dyw_i)\, .
\end{align*}
We have the following estimate 
\begin{align*}
    |A_i| &\le \sum_{j=1}^N |a_{ij}| e^{-(\yw_j-\yw_i)} \left| \cos\left(\xw_j-\xw_i\right) \right| \left| \dxw_j-\dxw_i \right| \\
    &\le e^{2\varepsilon} \sum_{j=1}^N |a_{ij}| \left(|\dxw_j|+ |\dxw_i| \right) \le 2e^{4\varepsilon}\Delta \sum_{j=1}^N |a_{ij}| \le 2e^{4\varepsilon}\Delta^2,
\end{align*}
and, similarly,
\begin{align*}
    |B_i| \leq 2e^{4\varepsilon}\Delta^2\, .
\end{align*}
Therefore,
\begin{align*}
    \left|\ddxw_i\right| &= |A_i-B_i| \le |A_i|+|B_i| \le 4e^{4\varepsilon}\Delta^2\, .
\end{align*}
\end{proof}

Combining the results established in Propositions \ref{prop:bound_theta} and \ref{prop:bound_x_t_w}, we are now ready to give a proof for \Cref{thm:main-error-1}.

\begin{proof} [Proof of \Cref{thm:main-error-1}] 
By Taylor's theorem, there exist $0<s_1, s_2<t$ such that 
\begin{align*}
    \theta^{\rm w}_i (t) &= \theta^{\rm w}_i (0) + \frac{t}{1!}\dot{\theta}^{\rm w}_{i} (0) + \frac{t^2}{2!} \ddot{\theta}^{\rm w}_i(s_1)\, ,
\end{align*}
and
\begin{align*}
    \xw_i(t) = \xw_i(0) + \frac{t}{1!} \dxw_i(0) + \frac{t^2}{2!} \ddxw_i(s_2)\, . 
\end{align*}
Since the initial conditions for $\bm{\theta}^{\rm w}$ and $\bmxw$ are chosen to be the same, we have that the zeroth-order terms in the two Taylor expansions agree, i.e., $\theta^{\rm w}_i (0) = \xw_i(0)$.
Moreover, noticing that the initial condition for ${\bm y}^{\rm w}$ is zero and comparing \Cref{eq:original_KM,eq:complex_Cartesian}, we have that the first-order terms agree as well, i.e., 
$\dot{\theta}^{\rm w}_{i} (0) = \dxw_i(0)$.
Therefore, we have
\begin{align*}
    \theta^{\rm w}_i (t) - \xw_i (t) = \frac{1}{2} \left[\ddot{\theta}^{\rm w}_i (s_1) - \ddxw_{i} (s_2) \right]t^2\, .
\end{align*}
By Propositions \ref{prop:bound_theta} and \ref{prop:bound_x_t_w}, we conclude that 
\begin{align*}
    \left|\theta^{\rm w}_i (t) - \xw_i (t)\right| &\le \frac{t^2}{2} \left(\left|\ddot{\theta}^{\rm w}_i(s_1)\right|+\left|\ddxw_i(s_2)\right|\right) 
    \le \frac{t^2}{2}\left(2\Delta^2+4e^{4\varepsilon}\Delta^2\right)
    =\left(1+2e^{4\varepsilon}\right)\Delta^2 t^2\, .
\end{align*}
Thus, \Cref{eq:local_error_bound} holds with $C_A = \left(1+2e^{4\varepsilon}\right)\Delta^2$.
\end{proof}

\begin{expl}
In \Cref{fig:one_window}, we illustrate the local error over one window for different values of $\tau$. We use the same graph and initial values as in \Cref{fig:kuramoto-comparison}. Here, $\tau$ ranges from $0.1$ to $1$. We observe that, in this case, the error is quadratic with respect to $\tau$, which is consistent with \Cref{thm:main-error-1}.
\begin{figure}[htbp]
 \centering
 \includegraphics[width=0.6 \textwidth ]{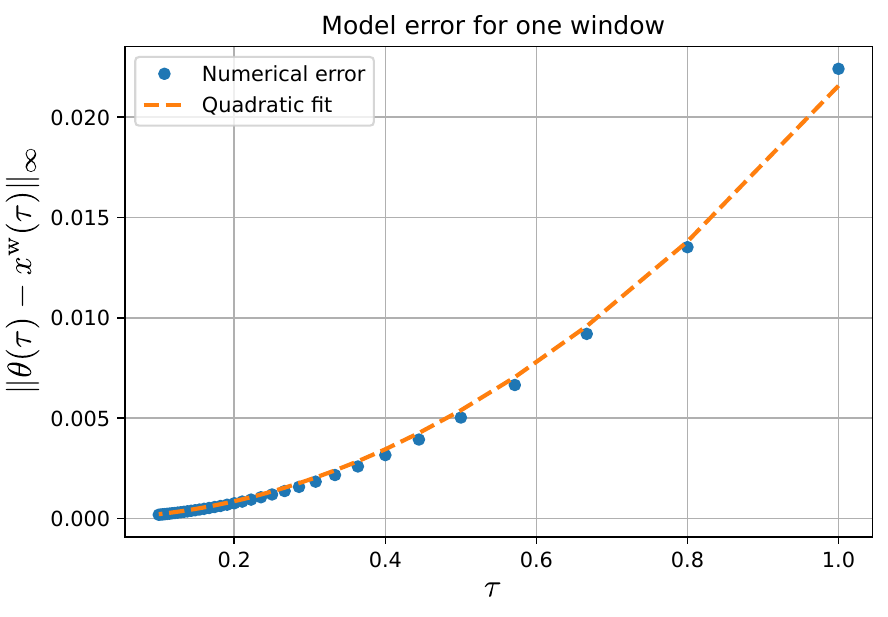}
\caption{Local error over one window for different values of $\tau$, using the same graph and initial values as in \Cref{fig:kuramoto-comparison}.}
\label{fig:one_window}
 \end{figure}
\end{expl}

\subsection{Initial-state error}
\label{sec:initial_state_error}
We analyze the initial-state error from \Cref{eq:error_split} in the following.
Let us compare the dynamics of the original Kuramoto model with two different initial conditions, expressed as
\begin{align*}
    \dot{\theta}_i = \sum_{j=1}^N a_{ij} \sin(\theta_j- \theta_i), \enskip \bm\theta(0)= \bm\theta_0, 
\end{align*}
and 
\begin{align*}
    \dot{\varphi}_i = \sum_{j=1}^N a_{ij} \sin(\varphi_j- \varphi_i), \enskip \bm\varphi(0)= \bm\varphi_0\, , 
\end{align*}
respectively.
Denoting the solutions of these two ODE systems by $\bm\theta(t)$ and $\bm\varphi(t)$, respectively, we have the following result:
\begin{thm}\label{thm:stability_error}
Let $\Delta$ be as defined in (\ref{eq:Delta_infinity}). We have
$
\|\bm\theta(t)-\bm\varphi(t)\|_\infty
\leq
e^{2\Delta t}\|\bm\theta_0-\bm\varphi_0\|_\infty$ for all $t \geq 0.$
\end{thm}

\begin{proof}
Let $\bm\psi(t) := \bm\theta(t) - \bm\varphi(t)$. 
Using the integral form of the two equations and the inequality
\begin{align*}
    |\sin (x)-\sin (y)|\leq |x-y|,
\end{align*}
we obtain, for each $1\leq i\leq N$,
\begin{align*}
    |\psi_i(t)| &= \left|\psi_i(0)+\int_0^t\sum_{j=1}^N a_{ij}\left[\sin(\theta_j(s)-\theta_i(s))-\sin(\varphi_j(s)-\varphi_i(s))\right] \,ds\right|\\
    & \leq |\psi_i(0)| + \int_{0}^t \sum_{j=1}^N |a_{ij}| \cdot |\sin(\theta_j(s)-\theta_i(s))- \sin(\varphi_j(s)-\varphi_i(s))| \, ds \\ 
    &\leq |\psi_i(0)| + \int_0^t \sum_{j=1}^N |a_{ij}| \left| (\theta_j(s)-\theta_i(s))- (\varphi_j(s)-\varphi_i(s)) \right| \,ds\\
    &\leq |\psi_i(0)| + \int_0^t \sum_{j=1}^N |a_{ij}| \bigl( |\psi_j(s)|+|\psi_i(s)| \bigr) \,ds\\
    &\leq \|\bm\psi(0)\|_\infty + 2\Delta \int_0^t \|\bm\psi(s)\|_\infty \,ds.
\end{align*}
Taking the maximum over $1\leq i\leq N$ gives
\begin{align*}
    \|\bm\psi(t)\|_\infty
    \leq
    \|\bm\psi(0)\|_\infty
    +   
    2\Delta
    \int_0^t
    \|\bm\psi(s)\|_\infty\,ds.
\end{align*}
Therefore, by Gr\"onwall's inequality,~\cite{Gronwall1919}  
\begin{align*}
    \|\bm\psi(t)\|_\infty
    \leq
    e^{2\Delta t}\|\bm\psi(0)\|_\infty = e^{2\Delta t}\|\bm\theta_0-\bm\varphi_0\|_\infty.
\end{align*}
\end{proof}

\subsection{Global error}
With the results established above, we are now ready to prove the main result stated in \Cref{thm:main} regarding the global error $E_k$.
Recall that the total observation time $T$ is fixed and divided into $n$ windows of equal size $\tau$.
We assume that $\tau \leq \tau_0$, where $\tau_0$ is given in \Cref{eq:tau_0}.
In the proof below, we establish that there exists a constant $C>0$, depending only on $T$ and $A$, such that $E_k \leq C\tau$ holds for $k = 1, \ldots, n$.

\begin{proof}[Proof of \Cref{thm:main}]
Recall the procedure of the windowed approach described in \Cref{sec:windowed_approach} and the error split in \Cref{eq:error_split}, where the global error $E_k$ is bounded by the sum of the local error and the initial-state error.
Since we assume $\tau \leq \tau_0$, the local error is bounded by $C_A \tau^2$ following \Cref{thm:main-error-1}, where $C_A$ depends only on $\Delta$ from \Cref{eq:Delta_infinity}, which in turn depends only on $A$. 
Applying \Cref{thm:stability_error} to the $k$th window and noticing that the difference between the two initial conditions satisfies $\left\|\bm{x}^\wk{k-1}(\tau) - \bm\theta((k-1)\tau)\right\|_\infty = E_{k-1}$, we have that the initial-state error is bounded by $e^{2\Delta\tau} E_{k-1}$.
Combining these two error bounds, we have the recursive relation 
\begin{equation}\label{eq:E_relation}
    E_k \leq C_A\tau^2 + e^{2\Delta\tau}E_{k-1}.
\end{equation}
It is obvious that $E_0 = 0$ since the windowed approach starts with the same initial condition as the original model. 
Iteratively applying \Cref{eq:E_relation} gives
\begin{equation}
    E_k 
    \leq 
    C_A\tau^2\sum_{\ell=0}^{k-1}e^{2\Delta\tau\ell}
    = 
    C_A\tau^2\,\frac{e^{2\Delta\tau k}-1}{e^{2\Delta\tau}-1}.    
\end{equation}
Because \(k\tau\leq n\tau \leq T\), we have $e^{2\Delta\tau k}-1 \leq e^{2\Delta T}-1\, ;$ moreover, since \(e^x-1\geq x,\, \forall x\geq0\), we have $e^{2\Delta\tau}-1\geq2\Delta\tau$.
Consequently,
\begin{equation}\label{eq:global_error_bound}
    E_k
    \leq
    C_A\tau^2
    \frac{e^{2\Delta T}-1}{2\Delta\tau} = 
    \frac{C_A}{2\Delta}
    \left(e^{2\Delta T}-1\right)\tau.
\end{equation}
Thus, we arrive at $E_k \leq C \tau$ with $C= \dfrac{C_A}{2\Delta} \left(e^{2\Delta T}-1\right)$, which depends on $T$ and $A$ only. 
\end{proof}

\begin{rem}
From \Cref{eq:global_error_bound}, we observe that for a fixed window size $\tau$, the global error $E_k$ is bounded by an exponential function of $T$, the total observation time.
This is consistent with various numerical methods for ODEs, such as the Euler method and the Runge--Kutta methods.\cite{atkinson2009numerical}
\end{rem}

\begin{expl}
In \Cref{fig:global_error}, we illustrate the global error for different values of $\tau$. We use the same graph and initial values as in \Cref{fig:kuramoto-comparison}. As in the local error experiment discussed in \Cref{fig:one_window}, $\tau$ ranges from $0.1$ to $1$. We observe that, in this case, the error is linear with respect to $\tau$, which is consistent with our analysis above.
\begin{figure}[htbp]
    \centering
    \includegraphics[width=0.6 \textwidth ]{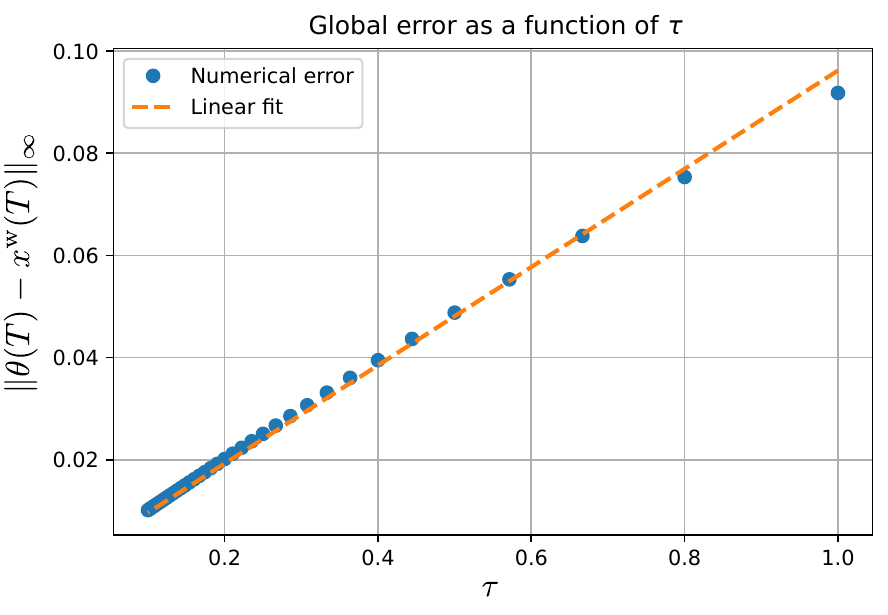}
    \caption{Global error as $\tau$ ranges from $0.1$ to $1$, using the same graph and initial values as in \Cref{fig:kuramoto-comparison}.}
    \label{fig:global_error}
\end{figure}

\end{expl}

\section*{Acknowledgments}
This work was completed while the third-named author (TTN) was a visiting faculty member at Argonne National Laboratory. He is grateful to Argonne for providing a wonderful and supportive working environment. 
He is also grateful to Dr. Nguyen Trung Hieu, who shared some insights on numerical ODE solvers and in particular  on Gr\"onwall's inequality. 
We acknowledge the use of AI to assist with writing the Python code used in our numerical experiments. 
All mathematical results were obtained by the authors. 
This research used resources of the Argonne Leadership Computing Facility, a U.S. Department of Energy (DOE) Office of Science user facility at Argonne National Laboratory and is based on research supported by the U.S. DOE Office of Science-Advanced Scientific Computing Research Program, under Contract No. DE-AC02-06CH11357.
RD was supported by the Swiss National Science Foundation under grand nr. 200021\_215336.

\appendix

\section{Error bound for the order parameter}\label{app:estimate}

We can use the global error $E_k$ to control the
difference between the two order parameters at the end
of the $k$th window. In fact, we have the following
\begin{prop}
For each $k$
    \begin{align}
\left|R_k-R^{\rm w}_k\right|
&\leq \frac{1}{N}\sum_{i=1}^N
\left|\theta_i(k\tau)-x^\wk{k}_i(\tau)\right|
\leq \left\|\bm{\theta}(k\tau)-\bm{x}^\wk{k}(\tau)\right\|_{\infty}
= E_k.
\end{align}
\end{prop}
\begin{proof}

By the definitions in
\Cref{eq:order_parameter_R,eq:order_parameter_R_complex_KM}
and the triangle inequality, we have
\begin{align}
\left|R_k-R^{\rm w}_k\right|
&= \frac{1}{N}\left|\bigg|\sum_{i=1}^N e^{\ci\theta_i(k\tau)}\bigg|
- \bigg|\sum_{i=1}^N e^{\ci x^\wk{k}_i(\tau)}\bigg|\right| \nonumber\\
&\leq \frac{1}{N}\left|\sum_{i=1}^N e^{\ci\theta_i(k\tau)}
- \sum_{i=1}^N e^{\ci x^\wk{k}_i(\tau)}\right|
\leq \frac{1}{N}\sum_{i=1}^N
\left|e^{\ci\theta_i(k\tau)} - e^{\ci x^\wk{k}_i(\tau)}\right|.
\label{eq:bound_order_parameter}
\end{align}

For any $a,b\in\mathbb{R}$, we can write
\begin{align}
    e^{\ci a}-e^{\ci b} &= e^{\ci(a+b)/2}\left(e^{\ci(a-b)/2}-e^{-\ci(a-b)/2}\right) = e^{\ci(a+b)/2} \cdot 2\ci \sin\left(\frac{a-b}{2}\right).
\end{align}
Since $\left|e^{\ci(a+b)/2}\right|=1$, it follows that
$
    \left|e^{\ci a}-e^{\ci b}\right| = 2\left|\sin\left(\frac{a-b}{2}\right)\right|\, .
$
Furthermore, since $ |\sin (x)|\leq |x|$, $\forall x \in \mathbb R$, we have 
\begin{align}
    \left|e^{\ci a}-e^{\ci b}\right| \leq 2\left|\frac{a-b}{2}\right| = |a-b|\, .
\end{align}

Applying this elementary result to \Cref{eq:bound_order_parameter}, we arrive at 
\begin{align}
\left|R_k-R^{\rm w}_k\right|
&\leq \frac{1}{N}\sum_{i=1}^N
\left|\theta_i(k\tau)-x^\wk{k}_i(\tau)\right|
\leq \left\|\bm{\theta}(k\tau)-\bm{x}^\wk{k}(\tau)\right\|_{\infty}
= E_k.
\end{align}
\end{proof}

\end{document}